\documentclass[10pt]{article}
\usepackage{graphicx}
\usepackage{amsmath}     
\usepackage{amssymb}    
\usepackage{epsfig}  
\def \BE {\begin{equation}}
\def \EE {\end{equation}}
\def \BEA {\begin{eqnarray}}
\def \EEA {\end{eqnarray}}

\def \one {^{(1)}}
\def \two {^{(2)}}

\begin{document}
\title{Wave turbulence and vortices in  Bose-Einstein condensation}
\author{Sergey Nazarenko$^*$ and  Miguel Onorato$^\dagger$\\
\\
$^*$ Mathematics Institute,
The University of Warwick, 
\\
Coventry, CV4-7AL, UK
\\
$^\dagger$Dipartimento di Fisica Generale,
 Universit\`a di Torino,\\
  Via P. Giuria, 1 - 10125 -
\\
Torino, Italy
}

\maketitle

\abstract{We report a numerical study of turbulence
and Bose-Einstein condensation within the two-dimmensional
Gross-Pitaevski 
model with repulsive interaction. In presence of weak 
 forcing localized around
some wave number in the Fourier space, we observe three qualitatively different
evolution stages. 
At the initial stage a thermodynamic energy equipartition spectrum   
forms at  both smaller and larger scales with respect to the forcing scale.
This agrees with predictions of the 
the four-wave kinetic equation of the Wave Turbulence (WT) theory.
At the second stage, WT breaks down at large scales and the interactions
become strongly nonlinear. Here, we 
observe formation of a gas of quantum vortices 
 whose number decreases due to an annihilation
process helped by the acoustic component. This process leads to formation of a
 coherent-phase  Bose-Einstein condensate.
After such a coherent-phase condensate forms, evolution enters a third
stage characterised by three-wave interactions of acoustic waves that can be described again 
using the WT theory. 
}

\section{Background and motivation}

For dilute gases with large energy occupation numbers 
the Bose-Einstein condensation (BEC) can be described by the
Gross-Pitaevsky (GP) equation \cite{Gross,Pitaevsky1961}:
\begin{equation}
 i \Psi_t + \Delta \Psi - |\Psi|^2 \Psi =  \gamma,
\label{gp}
\end{equation}
where $\Psi$ is the condensate ``wave function'' (i.e. the c-number
part of the boson annihilation field)
 and $\gamma$ is an operator which
models possible forcing and dissipation mechanisms which will be discussed
later. Renewed interest to the nonlinear dynamics
described by GP equation is related to relatively recent experimental discoveries of
BEC  \cite{Anderson,Bradley,Davis}. GP equation also describes light
behaviour in media with Kerr nonlinearities. In the nonlinear optics
context it is usually called the Nonlinear Schoedinger (NLS) equation.

It is presently understood, in both the nonlinear optics and BEC contexts,
that the nonlinear dynamics described by GP equation is typically
chaotic and often non-equilibrium \cite{SV,ZMR85,tkachev,DNPZ}.
Thus, it is best characterised as ``turbulence'' emphasizing its
resemblance to the classical Navier-Stokes (NS) turbulence.
On the other hand, the GP model has an advantage over NS
 because it has a weakly nonlinear limit in which the stochastic
field evolution can be represented as a large set of weakly interactive
dispersive waves.  A systematic statistical closure is possible for such 
systems and the corresponding theory is called Wave Turbulence (WT)
\cite{ZLF}.
For small perturbations about the zero state in the GP model, WT closure predicts that
the main nonlinear process will be four-wave resonant interaction.
This closure was used in \cite{SV,tkachev,DNPZ} to describe the
initial stage of BEC. In the present paper we will examine this description numerically.
We report that our numerics agree with the predicted by WT spectra
at the initial evolution stage.

It was also theoretically predicted that the four-wave WT closure will eventually
fail due to emergence of a coherent condensate state which is uniform
in space \cite{DNPZ}. At a later stage the condensate is so strong that the
nonlinear dynamics can be represented as interactions of small perturbations
about the condensate state. Once again, one can use WT to describe such a system,
but now the leading process will be a three-wave interaction of acoustic-like
waves on the condensate background \cite{DNPZ}. 
Coupling of such acoustic turbulence to the condensate was considered in
\cite{zakhnaz} which allowed to derived the asymptotic law of the condensate growth.
However, this picture relies
on assumptions that the system will consist of a {\em uniform} condensate
and {\em small} perturbations.
Neither the condensate uniformity nor the smallness of perturbations
have ever been validated before.
In the present paper we will examine
whether it is true that the late stage of GP evolution can be represented as
a system of weakly nonlinear acoustic waves about strong quasi-uniform condensate.
By
examining the frequency-wave number Fourier transforms, we do observe 
 waves with frequency in agreement with the
Bogoliubov dispersion relation.
The width of the frequency spectrum is narrow enough for these waves
to be called weakly nonlinear. 

A big unresolved question in the theory of GP turbulence
has remained about the stage of transition 
from the four-wave to the tree-wave regimes. This stage is strongly nonlinear
and, therefore, cannot be described by WT. However, using 
direct numerical 
simulations of equation (\ref{gp}),  we show that the transitional state involves a gas of annihilating vortices.
When the number of vortices reduces so that the mean distance between the
vortices becomes greater than the vortex core radius (healing length) the
dynamics becomes strongly nonlinear. This corresponds to entering the Thomas-Fermi
regime when the mean nonlinearity is greater than the dispersive term
in the GP equation. The mean inter-vortex distance is a measure of the correlation
length of the phase of $\Psi$ and, therefore, the vortex
annihilation corresponds to creation of a coherent-phase condensate.
At this point, excitations with wavelengths in between of the vortex-core
radius and the inter-vortex distance behave as sound.
In this paper, we draw attention to the similarity of this transition process
to
the Kibble-Zurek mechanism of the second-order phase transition
in cosmology \cite{kibble,zurek}.


\section{WT closure and predictions}

The WT closure is based on the assumptions of small nonlinearity and
of random phase and amplitude variables. To show how these assumptions come
about
we reproduce several essential steps of the kinetic equation derivation.
Let us consider GP equation (\ref{gp}) in a periodic box and write it
in the Fourier space:
 \begin{equation}
 i \partial_t \hat \Psi_k -k^2 \hat \Psi_k = 
\sum_{\alpha , \mu, \nu } 
\bar{\hat  \Psi}_\alpha \hat \Psi_\mu \hat \Psi_\nu \delta^{k\alpha}_{\mu\nu}
 + \hat \gamma_k,
\label{k_gp}
\end{equation}
where $\hat \Psi_j = \hat \Psi({\bf k}_j)$, overbar means complex conjugation,
 wave vectors ${\bf k}_j \; (j=1,2,3)$ are on
a 2D grid (due to periodicity) and
symbol $\delta^{k\alpha}_{\mu\nu} =1$ for ${\bf k} + {\bf k}_\alpha = {\bf k}_\mu +
{\bf k}_\nu$  and equal to 0 otherwise.

\subsection{Four-wave interaction regime}
In order to describe the WT theory for equation (\ref{k_gp}) we will 
neglect the forcing/dissipation
term $ \hat \gamma_k$ assuming that those are localized at high or low
wave numbers and we are mainly 
interested in an inertial range of ${\bf k}$.
Let us make a transformation to the interaction representation variables $a_k$
which absorb the linear dynamics and oscillations due to the diagonal terms
${\bf k}={\bf k}_2$ and ${\bf k}_1 ={\bf k}_3$,
\begin{equation}
a_k = e^{i \omega_k t + i \int \omega_{nl} \, dt} \Psi_k,
\label{ak}
\end{equation}
where $ \omega_k = k^2 $ is the linear frequency and
\begin{equation}
\omega_{nl} = \sum_{\bf k} |\hat \Psi_k|^2
\label{omegaNL}
\end{equation}
is the nonlinear frequency shift. In terms of $a_k$ we have
 \begin{equation}
 i \partial_t a_k =
\sum_{\alpha \ne \mu, \nu }
\bar a_\alpha  a_\mu  a_\nu \,\delta^{k\alpha}_{\mu\nu} \,
 e^{i \omega^{k\alpha}_{\mu \nu} \, t},
\label{k_gp_a}
\end{equation}
where $\omega^{k\alpha}_{\mu \nu} = \omega_k + \omega_\alpha
- \omega_\mu -\omega_\nu$. The nonlinear frequency drops
out from this expression because it is $k$-independent.
Small nonlinearity allows to separate the linear and nonlinear
timescales.
Let us consider solution $a_k(T)$ at an intermediate (between
linear and nonlinear) time $T \gg 2 \pi / \omega_k$ 
but such that changes in $a_k$ are still small. We seek this solution
 as a series
 \begin{equation}
a_k = a_k^{(0)} + a_k^{(1)} + a_k^{(2)} + ...
\label{a_series}
\end{equation}
which is obtained by recursive substitution into
(\ref{k_gp_a}). For example, $a_k^{(0)} = a_k|_{T=0}$ and
\begin{eqnarray}
a_k\one(T) = - i \sum_{\alpha\mu\nu}
 \bar a_\alpha^{(0)} a_\mu^{(0)} a_\nu^{(0)} \delta^{k\alpha}_{\mu\nu} \Delta^{k
 \alpha}_{\mu\nu}.  \label{FirstIterate}
\end{eqnarray}
where
 $\Delta^{l \alpha}_{\mu\nu} =
({e^{i\omega^{l\alpha}_{\mu\nu}T}-1})/{i \omega^{l\alpha}_{\mu\nu}}.
 \label{NewellsDelta}$
For the second iteration we get
\begin{eqnarray}
a_k\two(T)=\sum_{\alpha \ne \mu, \nu; u \ne v, \beta}\left( 
\delta^{\mu\nu}_{\alpha u} \delta^{ku}_{v\beta}
 a_\alpha^{(0)} a_v^{(0)} a_\beta^{(0)} 
\bar a_{\mu}^{(0)} \bar a_{\nu}^{(0)} 
E(\omega^{k\mu\nu}_{\alpha v \beta}, \omega^{ku}_{v\beta}) 
- 
2 \delta^{\alpha v}_{\mu\nu} \delta^{k u}_{v \beta}
\bar a_{\alpha}^{(0)} \bar a_{u}^{(0)}
a_{\mu}^{(0)} a_{\nu}^{(0)} a_\beta^{(0)} 
E( \omega^{k\alpha u}_{\mu\nu\beta},  \omega^{ku}_{v\beta})\right) ,
\label{SecondIterate}
\end{eqnarray}
where
$E(x,y)=\int_0^T \Delta(x-y)e^{i y t} d t .$
Expressions (\ref{FirstIterate})  and  (\ref{SecondIterate}) are sufficient
for
writing the leading order (in small nonlinearity) for the evolution
of the spectrum, $n_k = \langle |a_k|^2 \rangle$:
 \begin{eqnarray}
\dot n_k = (\langle |a_k(T)|^2 \rangle - \langle |a_k(0)|^2 \rangle)/T = 
{1 \over T} (\langle a_{k}^{(0)} \bar  a_{k}^{(1)} \rangle + 
\langle a_{k}^{(1)} \bar  a_{k}^{(0)} \rangle +
\langle a_{k}^{(0)} \bar  a_{k}^{(2)} \rangle + 
\langle a_{k}^{(2)} \bar  a_{k}^{(0)} \rangle +
\langle |a_{k}^{(1)}|^2  \rangle).
\label{ndot}
\end{eqnarray}
Here, we must substitute expressions (\ref{FirstIterate})  and
(\ref{SecondIterate}) and perform averaging over the ensemble of
initial fields $a_k^{(0)} $. For this, we will use a generalized
RPA (Random Phase and Amplitude) approach introduced in \cite{cln}
and which is different from the traditional RPA (Random Phase approximation,
see e.g. \cite{ZLF})
by emphasizing randomness of the amplitudes and not only the phases.
Namely, we represent complex amplitudes as $a_k^{(0)} = A_k \psi_k$ where
$A_k $  are positive real amplitudes and
$\psi_k$  are phase factors taking values on the unit circle on the complex
plane. Generalized RPA says that the initial wave field is such that all
of its amplitudes $A_k $ and phase factors $\psi_k$ make a set of independent
random variables and that $\psi_k$'s are uniformly distributed on the unit
complex circle (distribution of $A_k $'s need not be specified).

RPA allows to close equations for the spectrum (\ref{ndot}) by using the 
Wick-type splitting of the higher Fourier moments in terms of the
spectrum.
The first two terms in (\ref{ndot}) contain three $a^{(0)} $'s each and,
therefore, vanish due to the phase randomness. The other three terms, after
substitution of (\ref{FirstIterate})  and
(\ref{SecondIterate}), RPA averaging and taking the large-box and large $T$ 
limits, give (see details e.g. in \cite{cln}):
\begin{equation}
\dot n_k = 4 \pi \int 
n_k n_u n_\mu n_\nu \left( { 1 \over n_k} + { 1 \over n_u}
-{ 1 \over n_\mu} - { 1 \over n_\nu} \right)
\delta(\omega^{ku}_{ \mu \nu})   \delta^{ku}_{ \mu \nu} 
 \, d{\bf k}_u  d{\bf k}_\mu  d{\bf k}_\nu.
\label{wke}
\end{equation}
This is the wave-kinetic equation (WKE) which is the most important
object in the wave turbulence theory (for the GE equation, it 
was first derived in \cite{ZMR85}). It contains Delta functions
for four wave vectors, $ \delta^{ku}_{ \mu \nu} =    
\delta({\bf k}+{\bf k}_u  -{\bf k}_\mu  -{\bf k}_\nu)$, and for
the four corresponding frequencies, $\delta(\omega^{ku}_{ \mu \nu}) $,
which means that the spectrum evolution in this case is driven by
a 4-wave resonance process.
Note that WT approach is applicable not only to the spectra but
also to the higher moments and even the probability density functions
\cite{cln,ln}. However, we are not going to reproduce these results
because their study is beyond the aims of the present paper.

There are four power-law solutions of the 4-wave
kinetic equation in (\ref{wke}) and they are related to the two invariants for such systems,
the total energy, $E = \int \omega_k n_k d{\bf k}$,
 and the total number of particles,
 $N = \int n_k d{\bf k}$.
Two of such power-law solutions correspond to a thermodynamic equipartition of one
of these invariants, 
\begin{eqnarray}
n_k \sim 1/\omega_k &=& k^{-2} \;\;\;\;\; \hbox{(energy equipartition)}, 
\label{n_e} \\
n_k \;\;\; &=& \hbox{const}  \;\;\; \hbox{(particle equipartition)}. 
\label{n_p}
\end{eqnarray}
These two solutions are limiting cases of the general thermodynamic
distribution,
\begin{equation}
n_k =  T/(\omega_k + \mu) ,
\end{equation}
where constants $T$ and $\mu$ have meanings of temperature and
chemical potential respectively.
Due to isotropy, it is convenient to deal with an angle-averaged
1D wave action density in variable $k = |{\bf k}|$, so called
1D wave action spectrum $N_k = 2 \pi k n_k$. In terms of $N_k$, solutions
(\ref{n_e}) and (\ref{n_p}) have exponents $-1$ and $1$ respectively.

The other two power-law solutions correspond to a Kolmogorov-like constant flux
of either energy (down-scale cascade) or the particles (up-scale
cascade) \cite{DNPZ}. As shown in \cite{DNPZ}, 
the formal solution for the inverse cascade has wrong sign of the
particle flux and is, therefore, irrelevant.
On the other hand, the power exponent of the 
direct cascade solution formally coincides with the energy equipartition
exponent $-2$ and, in fact, it is the same solution.
Because of such a coincidence, the energy flux value is equal to zero
on such a solution and, therefore, it is more appropriate to associate
it with thermodynamic equilibrium rather than a cascade.

\subsection{Three-wave interaction regime} \label{3-w-regime}
If the system is forced at large wave numbers and there is no dissipation
at low $k$'s then there will be condensation of particles at large scales.
The condensate growth will eventually lead to a breakdown of the
weak nonlinearity assumption \cite{DNPZ,biven} and the 4-wave WKE
(\ref{wke}) will become invalid for describing subsequent evolution.
On the other hand, it was argued in  \cite{DNPZ} that such late evolution
one can consider small disturbances of  coherent condensate
state $\Psi_0=$~const, so that a WT approach can be used again (but now
on a finite-amplitude background),
\begin{equation}
\Psi({\bf x},t) = \Psi_0 \, (1+ \phi({\bf x},t)), \;\;\;\; \phi \ll \Psi_0.
\label{psi_phi}
\end{equation}
Then, with respect to condensate perturbations $\phi$, the linear  dynamics
has to be diagonalised via the Bogoljubov transformation, which in our case is
\cite{DNPZ,lnw,zakhnaz}
 \begin{equation}
\hat \phi_k  = {1 \over 2 \sqrt{\rho_0}} \left[ \left({k \over \omega_k^{1/2}} +
{\omega_k^{1/2} \over k} \right) a_k +
 \left({k \over \omega_k^{1/2}} +
{\omega_k^{1/2} \over k} \right) \bar a_k \right],
\label{bogolubov}
\end{equation}
where $a_k$ are new normal amplitudes (see for example
\cite{zakhnaz}). In the linear approximation, amplitudes   $a_k$
oscillate at  frequency
\begin{equation}
\omega_k = k \sqrt{k^2 + 2 \rho_0 }
\label{bog_disp}
\end{equation}
which is called the Bogoljubov dispersion relation. For strong condensate,
$\rho_0 \gg k^2$, this dispersion relation corresponds to sound.

Because of the non-zero background, the nonlinearity will be quadratic
with respect to the condensate perturbations and, thus, the
resulting WT closure now gives rise to a 3-wave WKE. This WKE was first
obtained in \cite{DNPZ} (see also \cite{zakhnaz}) and here we reproduce
it without derivation,
\begin{eqnarray}
   \dot n_k = \pi \int (R_{k12} -R_{1k2} - R_{2k1})\,
                      d {\bf k}_{1} d {\bf k}_2 \, , 
   \label{KE3wave}
\end{eqnarray}
where 
$$ R_{k12} = |V_{k k_1 k_2}|^2 \, 
 \delta({{{\bf k} - \bf{k_1}-\bf{k_2}}}) \, 
           \delta ({\omega_{{ k}} -\omega_1 -\omega_2} ) \,
(n_1 n_2 - n_k n_1 - n_k n_2). $$
Here,
 $ V_{ k, k_1, k_2}$ is the interaction coefficient which has the
following form
\begin{equation}
 V_{ k, k_1, k_2} = {\Psi_0 \sqrt{ \omega_{ k} \omega_1 \omega_2}
\over  (2 \pi)^{3/2} } \left\{ {6  \over (\alpha_{k} \alpha_1 
\alpha_2)^{1/2}} + {1 \over 2}  \left[ { ({\bf k} \cdot {\bf k_1}) \over k k_1 
\alpha_2 } + { ({\bf k} \cdot {\bf k_2}) \over k k_2 
\alpha_1 } + { ({\bf k_1} \cdot {\bf k_2}) \over k_1 k_2 
\alpha_{k} } \right] \right\}
\end{equation}
where
\begin{equation}
\alpha_{ k} = {2 \rho_0 + k^2}.
\end{equation}
At late time the condensate becomes strong, $\rho_0 \gg k^2$, and turbulence
becomes of acoustic type. The number of particles is not conserved by
the turbulence alone (particles can be transferred to the condensate)
and there are only two relevant power-law solutions in this case:
thermodynamic equipartition of energy and the energy cascade spectrum.
Because of isotropy, one often considers 1D (i.e. angle-integrated) energy
density,
\begin{equation}
E(k) = 2 \pi k \omega_k n_k.
\end{equation}
In terms of this quantity, the thermodynamic spectrum is 
 \begin{equation}
E(k) \sim k,
\end{equation}
 and the energy cascade spectrum is
 \begin{equation}
E(k) \sim k^{-3/2}.
\end{equation}
Note that the energy cascade is direct and the corresponding spectrum
can be expected in $k$'s higher than the forcing wave number,
whereas the thermodynamic spectrum is expected at the low-$k$ range
to the left of the forcing \cite{zakhnaz}.

Note that the described above picture of acoustic WT relies on two
major assumptions.

1. Condensate is coherent enough so that its spatial variations
are slow and it can be treated as uniform when evolution of the perturbations 
about the condensate is considered. In the other words, a scale separation
between the condensate and the perturbations occurs.

2. Coherent condensate is much stronger than the chaotic acoustic
   disturbances. This allows to treat nonlinearity of the perturbations
around the condensate as small.

Both of these assumptions have not been validated before and their
numerical check will be one of our goals.
Another major goal will be to study the transition stage that lies in between of the 4-wave
and the 3-wave turbulence regimes. This transition is characterised by
strong nonlinearity and the role of numerical simulations becomes
crucially important in finding its mechanisms. 

Once the 3-wave acoustic regime has been reached, the condensate continues
to grow due to a continuing influx of particles from the acoustic turbulence
to the condensate. This evolution, where an unsteady condensate is coupled
with acoustic WT, was described in  \cite{zakhnaz} who predicted that
asymptotically the condensate grows as $\rho_0 \sim t^2$ if the forcing 
is of an instability type $\hat \gamma = \nu_k n_k$.
However, in the present paper we work with a different kind of forcing which
is most convenient and widely used in numerical simulations: we keep
amplitudes in the forcing range fixed (and we chose their phases randomly).
Thus, one should not expect observing the $ t^2$ regime predicted in
\cite{zakhnaz} in our simulations.
Note that 2D NLS turbulence was simulated numerically with
specific focus on the condensate growth rate in \cite{DYA96}.
In our work, we do not aim to study the condensate growth rate because
it is strongly dependent on the forcing type which, in our model, is
quite different from turbulence sources in laboratory.
On the other hand, we believe that the main stages of the condensation, i.e.
transition from a 4-wave process, through vortex annihilations, to
3-wave acoustic turbulence, are robust under a wide range of forcing types.


\section{Setup for numerical experiments}

In this paper we consider a setup corresponding to homogeneous
turbulence and, therefore, we ignore finite-size effects due
to magnetic trapping in BEC or to the finite beam radii in
optical experiments.
%
 For numerical simulations, we have used a standard  pseudo-spectral
method for the 2D equation (\ref{gp}). The number of grid
points in physical space was set to $N \times N$ with $N=256$. 
Resolution in Fourier space was $\Delta k=2\pi/N$.
Sink  at high wave numbers was provided by adding 
to the right hand side of  equation (\ref{gp}) 
the hyper-viscosity term  $\nu(-\nabla^2)^n \psi$.
Values of $\nu$ and $n$ were selected in order
to localized as much as possible
dissipation to high wave numbers but avoiding 
at the same time the bottleneck effect.
We have found, after a number of trials,
that $\nu=2 \times 10^{-6}$ and $n=8$ were
good choices for our purposes.
In some simulations,  we have also used a
dissipation at low wave numbers of the form of 
$\nu(-\nabla^2)^{-n} \psi$ with $\nu=1 \times 10^{-18}$
 and $n=8$.
This was done, e.g., to see what changes if one suppresses the condensate
formation.
 Forcing was localized in Fourier space and was chosen 
as $f=|f| exp[-i \phi(t)]$ with $|f|$ constant in time 
and $\phi(t)$ randomly selected between 0 and $2\pi$ each time step. 
i) To study turbulence in the down-scale inertial range we force the system
isotropically at wave numbers 
$4 \Delta k \leq |k| \leq  6 \Delta k$. To avoid condensation
at large scales we introduce 
a dissipation at low wave numbers, as was previously explained.
ii) To study the condensation we chose forcing at  wave numbers
$60 \Delta k \leq |k| \leq  63 \Delta k$ and dissipation 
at all higher wave numbers. A number of numerical simulations 
were  performed both with and without the dissipation 
at the low wave numbers.
Time step for integration was $t=0.1$ and usually $10^5$
time steps have been performed for each simulation. This
is usually enough for reaching a steady state when 
 dissipation at both high and low wave numbers was placed.

\section{Numerical results}
\subsection{Turbulence with suppressed condensation}

We start with a state without condensate for which WT predicts
four-wave interactions. WKE has two conserved quantities in this case,
the energy and the particles, and the directions of their transfer in the
scale space must be opposite to each other. 
Indeed, let us assume that energy flows up-scale and that it
gets dissipated at a scale much greater than the forcing
scale. This would imply dissipating the number of particles  
which is much greater than what was generated at the source (because of the factor $k^2$
difference between the energy and the particle spectral densities).
 This is impossible in steady state and, therefore, 
energy has to be dissipated at smaller (than forcing) scales.
 On the other hand, the
particles have to be transferred to larger scales because dissipating them
at very small scales would imply dissipating more energy than produced by
forcing. This speculation is standard for the systems with two positive
quadratic invariants, e.g. 2D Euler turbulence where one invariant, the energy,
flows up-scale and another one, the enstrophy, flows down-scale.

Thus, ideally, one would like to place forcing at an intermediate
scale and have two inertial ranges, up-scale and down-scale of the source.
However, this setup is unrealistic because the presently available
computing power would not allow us to achieve simultaneously two inertial
ranges wide enough to study scaling exponents. Therefore, we split this problem
to two, with forcing at the left and at the right ends of a single inertial
range.
 
\subsubsection{Turbulence down-scale of the forcing}

\begin{figure}
\centerline{\includegraphics[width=0.5\textwidth]{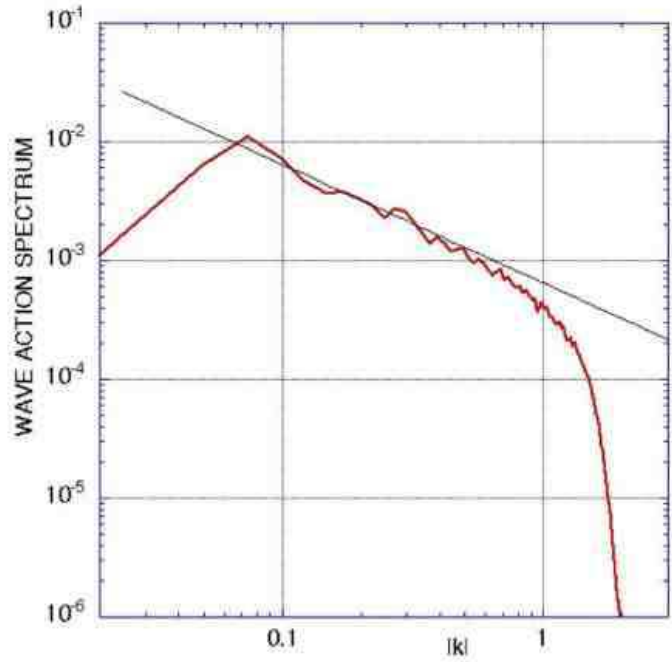}}
\caption{\label{fig: direct} 1D wave action spectrum $N_k$
for the down-scale inertial range.
A line corresponding to $k^{-1}$, the wave turbulence
prediction, is also included.}
\end{figure}

Our first numerical experiment is designed to test the WT
predictions about the turbulent state corresponding to the 
down-scale range with respect to the forcing scale. Thus we chose to force turbulence
at large scales and to dissipate it at the small scales as described
in the previous section. Our results for the one-dimensional wave action
 spectrum  is statistically stationary condition is shown in Figure \ref{fig: direct}. We see a range with 
slope $-1$ predicted by both the Kolmogorov-Zakharov (KZ) energy cascade
and the thermodynamic energy equipartition solutions of the 
four-wave WKE. As we mentioned earlier, it would be more appropriate
to interpret this spectrum as a quasi-thermodynamic state rather
than the KZ cascade because the energy flux expression formally turns into
zero at the power spectrum with $-1$ exponent. We emphasise, however, that
the state here is quasi-thermodynamic with a small flux component present
on thermal background because of the presence of the source and sink.
One could compare this state to a lake with two rivers bringing the water
in and out of the lake. In comparison, pure KZ cascade would be more
similar to a waterfall.

\begin{figure}
\centerline{\includegraphics[width=0.5\textwidth]{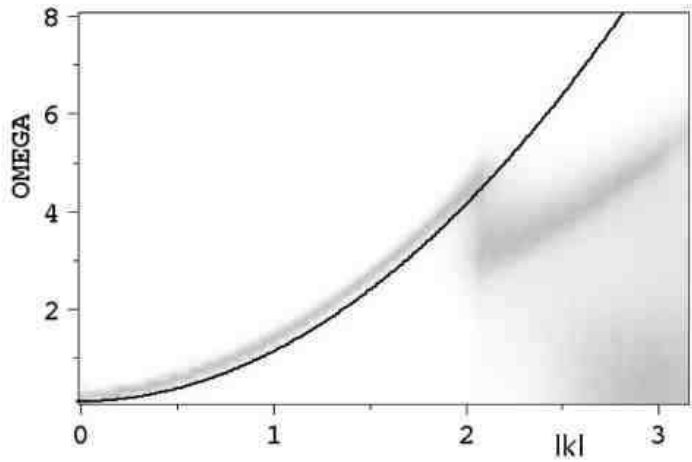}}
\caption{\label{fig:disp_rel_direct} 
Wave number-frequency distribution of the space-time Fourier transform
of $\Psi$ in the down-scale inertial range.
Dispersion relation from 
linear theory is shown in black curve.}
\end{figure}

To check that the waves in our system are indeed weakly nonlinear,
we look at the space-time Fourier transform of the wave field.
The frequency-wave number plot of this Fourier transform is shown in 
Figure \ref{fig:disp_rel_direct}. We see that this Fourier transform is
narrowly concentrated near the linear dispersion curve, which
confirms that the wave field is weakly nonlinear. We can also see that the spectrum is slightly shifted upwards by
a value which agrees with the nonlinear frequency shift
found via substitution of the numerically
obtained spectrum into (\ref{omegaNL}).

\subsubsection{Up-scale turbulence}
\begin{figure}
\centerline{\includegraphics[width=0.5\textwidth]{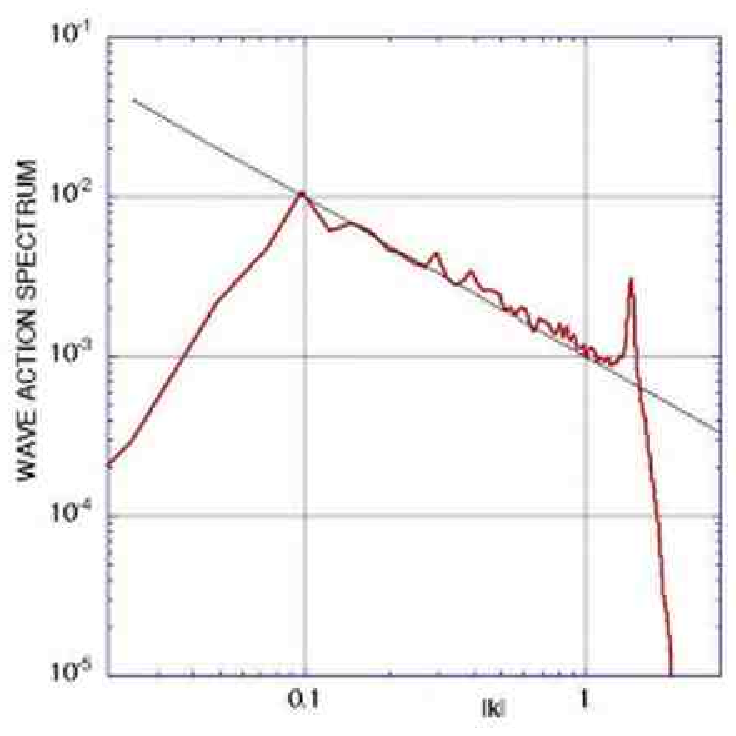}}
\caption{\label{fig: inverse} 1D waveaction spectrum $N_k$
 in the up-scale range. 
A power law of the form of $k^{-1}$ is also shown.}
\end{figure}

In the up-scale range  
one could expect that, in analogy with the 2D
Navier-Stokes turbulence, there would be an inverse cascade of the
number of particles  and that the corresponding KZ spectrum would be observed.
Nevertheless, it was pointed out in \cite{DNPZ}
that the analytical KZ spectrum has ``wrong'' direction 
of the flux of particles in the 2D
GP model and, therefore, cannot form.
Our numerics agree with this view.
Instead of the KZ, our numerical simulations show that 
a statistical stationary state with a power
 law very close to $k^{-1}$ forms,
see Figure  \ref{fig: inverse}. 
This solution 
corresponds to the thermodynamics solution with
energy equipartition in the $k$-space.
Note that both theoretical rejection of the
particle-cascade spectrum  \cite{DNPZ} and our numerical study relate
to the 2D model and the situation can change in the 3D case.
\footnote{
Another difference with the 3D case may be that in 3D the condensate
forms as a sharp peak at the ground state whereas in 2D
 no such peak is observed \cite{colm}. Such absence of a sharp peak at $k=0$
in 2D is consistent with our numerics.}
Namely, it is possible that the up-scale dynamics in 3D will be
characterised by the particle-flux KZ solution or a more
complicated mixed state which involves both cascade and temperature.
On the other hand, formation of a pure thermodynamic state in 2D
is quite fortunate for the theoretical description because
analogies with the theories of phase transition between
different types of thermodynamic equilibria become more meaningful.

\begin{figure}
\centerline{\includegraphics[width=0.5\textwidth]{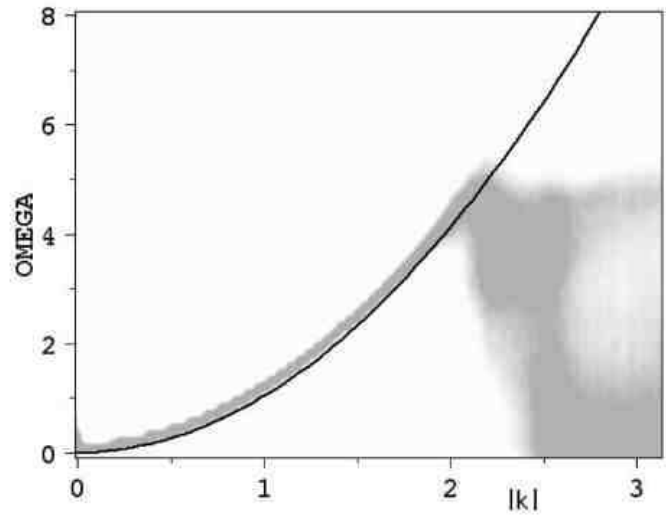}}
\caption{\label{fig:disp_rel_indirect} 
Wave number-frequency distribution of the space-time Fourier transform
of $\Psi$  in the up-scale inertial range.
Dispersion relation from 
linear theory is shown in black curve.}
\end{figure}

Here, we also check that the waves in this regime are weakly nonlinear
by looking at the space-time Fourier transform.
The corresponding frequency-wave number plot  is shown in Figure
\ref{fig:disp_rel_indirect}. As in the down-scale inertial range, we 
see that this Fourier transform is
narrowly concentrated near the linear dispersion curve, i.e. 
the wave field is weakly nonlinear in this state.
 
\subsection{Bose-Einstein condensation}
\subsubsection{Initial stage: four-wave process}

\begin{figure}
\centerline{\includegraphics[width=0.4\textwidth]{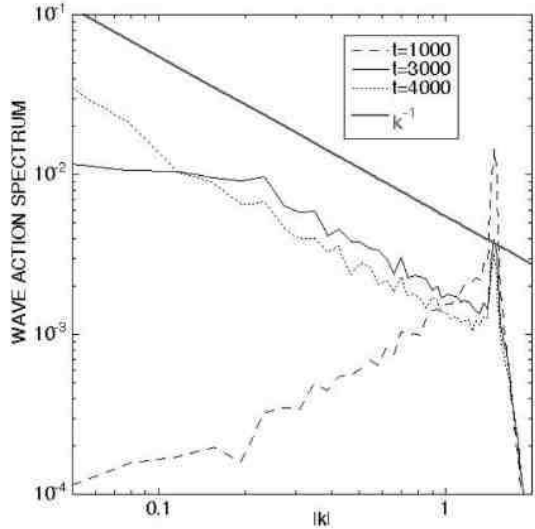}}
\caption{\label{fig: inverse01} Initial stages 
of the evolution of the 1D wave action spectrum $N_k$.
A power law of the form of $k^{-1}$ is also shown.}
\end{figure}

In order to study the stages of 
the condensation process, the results presented 
in the following  have been obtained with  
forcing localized at high wave numbers {\em without}
dissipation at low wave numbers.
 At the initial stage of the simulation, the nonlinearity
remains small compared to the dispersion in the GP equation and the
four-wave kinetic equation can be used. 
In Figure  \ref{fig: inverse01}, we show the initial 
(pre-condensate) stages of the spectrum  evolution. 
Similarly to the case where the condensation was
suppressed, we observe the formation of a thermodynamic 
distribution. 

\begin{figure}
\centerline{\includegraphics[width=0.4\textwidth]{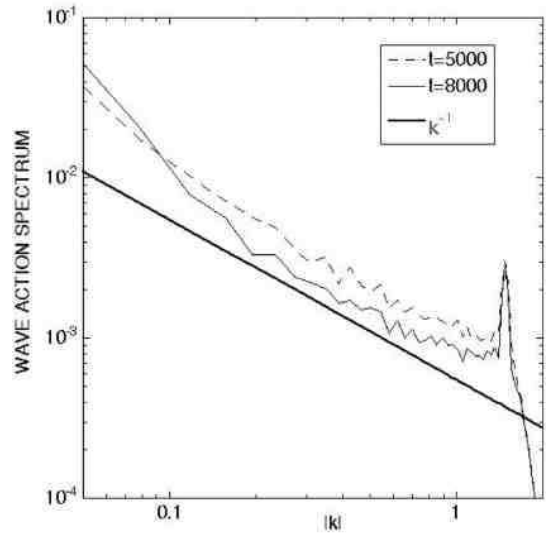}}
\caption{\label{fig: inverse02} Later stages 
of the evolution of the 1D wave action spectrum $N_k$.
A power law of the form of $k^{-1}$ is also shown.}
\end{figure}
\subsubsection{Transition}
After the stage where the four-wave interaction 
dynamics holds, the dynamics is characterised by a 
transitional stage in which the 
 the low-$k$ front of the evolving spectrum reaches the
largest scale (at about $t=5000$), see Figure 
\ref{fig: inverse02};
the spectrum 
begins to become steeper at low wave numbers and,
as expected, the thermodynamics solution does not
hold anymore.
This behaviour indicates that a change of regime occurs
around time $t=5000$.
However, the information contained in the spectrum is insufficient
to fully characterize this regime change and this brings us 
to study this phenomenon by measuring several
other important quantities.

\begin{figure}
\centerline{\includegraphics[width=0.4\textwidth]{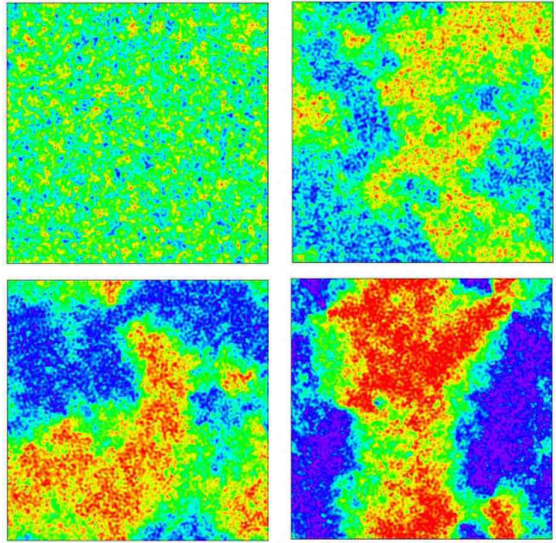}}
\caption{\label{fig: repsi}  $Re[\psi(x,y)]$
at different times: $t=2500$, $t=5000$, $t=7500$, $t=10000$.}
\end{figure}

To get an initial impression of what is happening during the transition
stage it is worth to  first of all examine the field distributions in the 
coordinate space. Figure \ref{fig: repsi}
 shows a series of frames of the real part of
$\Psi$ (imaginary part looks similar). One can see that this 
field exhibits growth of large-scale structure.

\begin{figure}
\centerline{\includegraphics[width=0.4\textwidth]{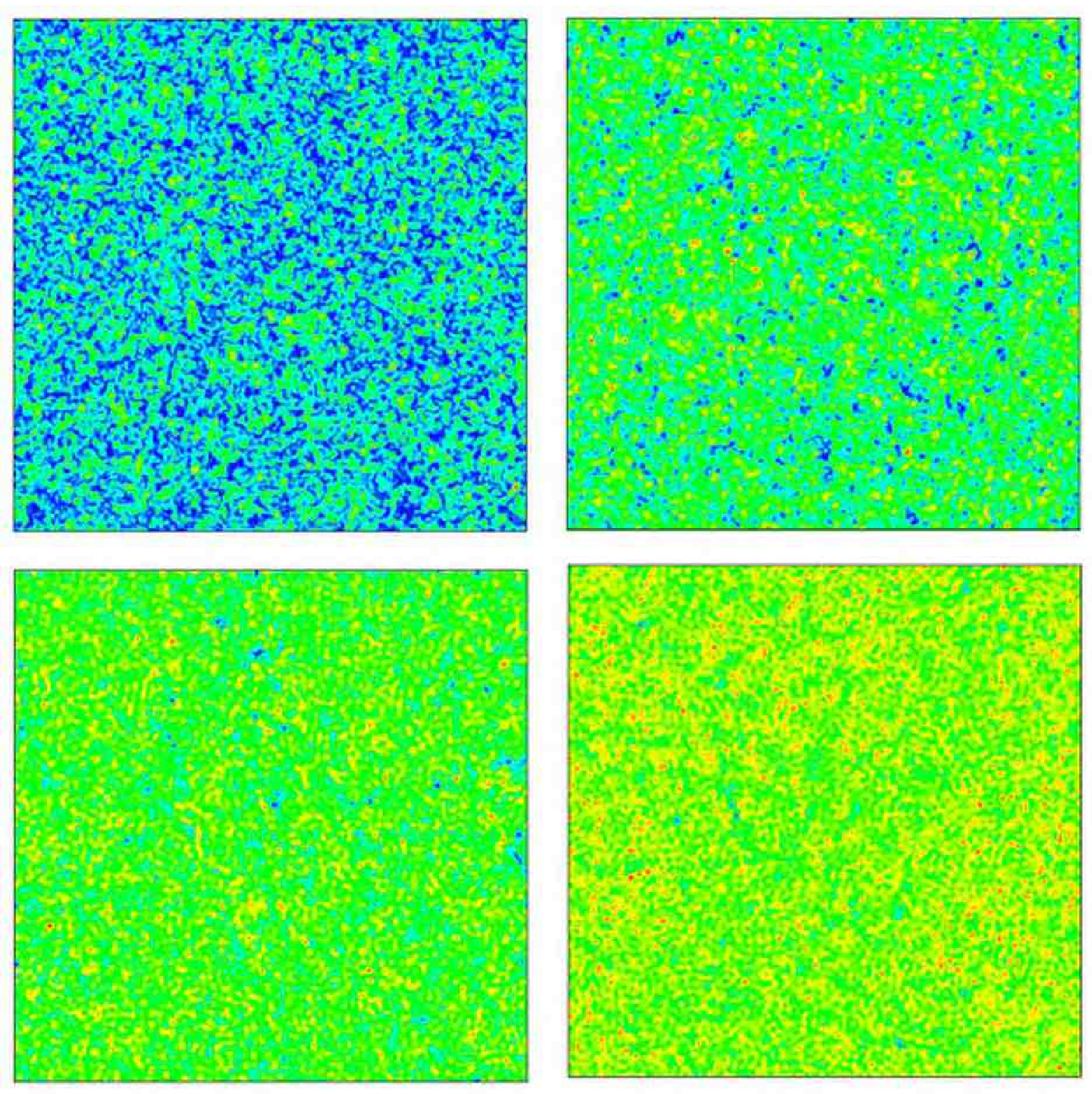}}
\caption{\label{fig: |psi|}  $|\psi(x,y)|$
at different times: $t=2500$, $t=5000$, $t=7500$, $t=10000$.}
\end{figure}

\begin{figure}
\centerline{\includegraphics[width=0.4\textwidth]{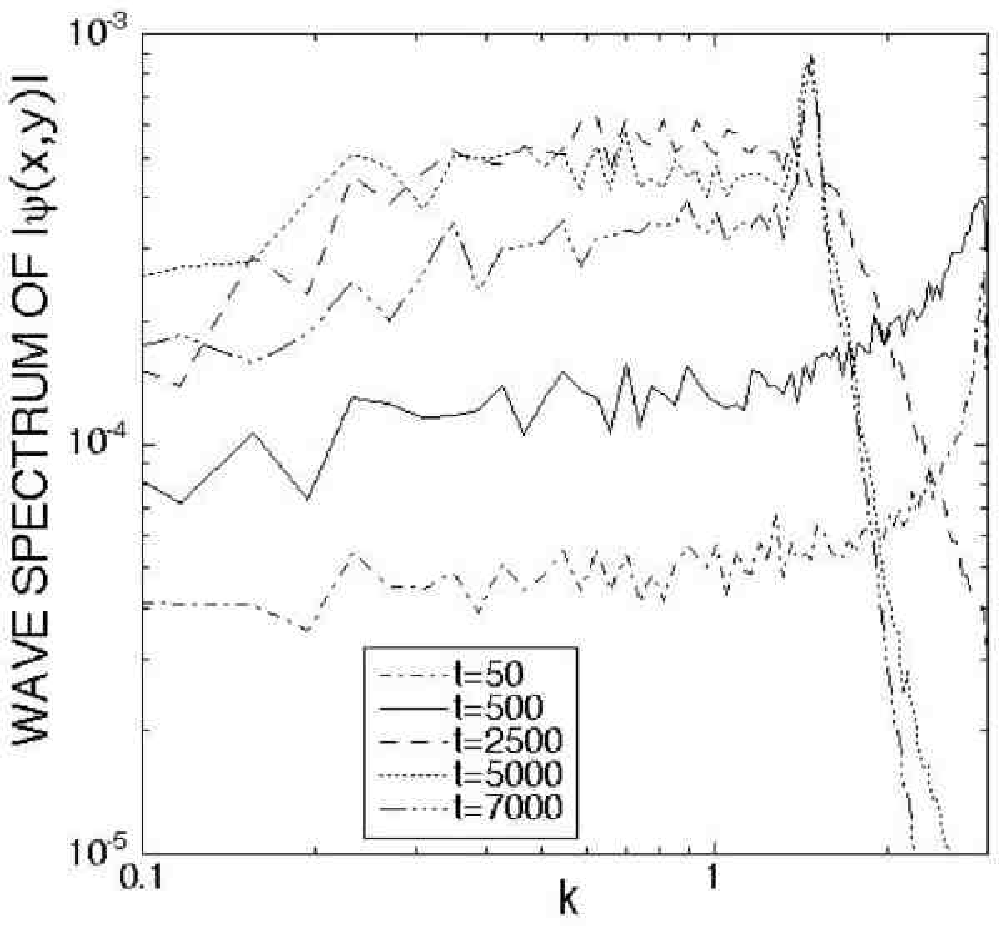}}
\caption{\label{fig: spettro|psi|} Spectrum for variable
$|\psi(x,y)|$ at  different times.}
\end{figure}

\begin{figure}
\centerline{\includegraphics[width=0.8\textwidth]{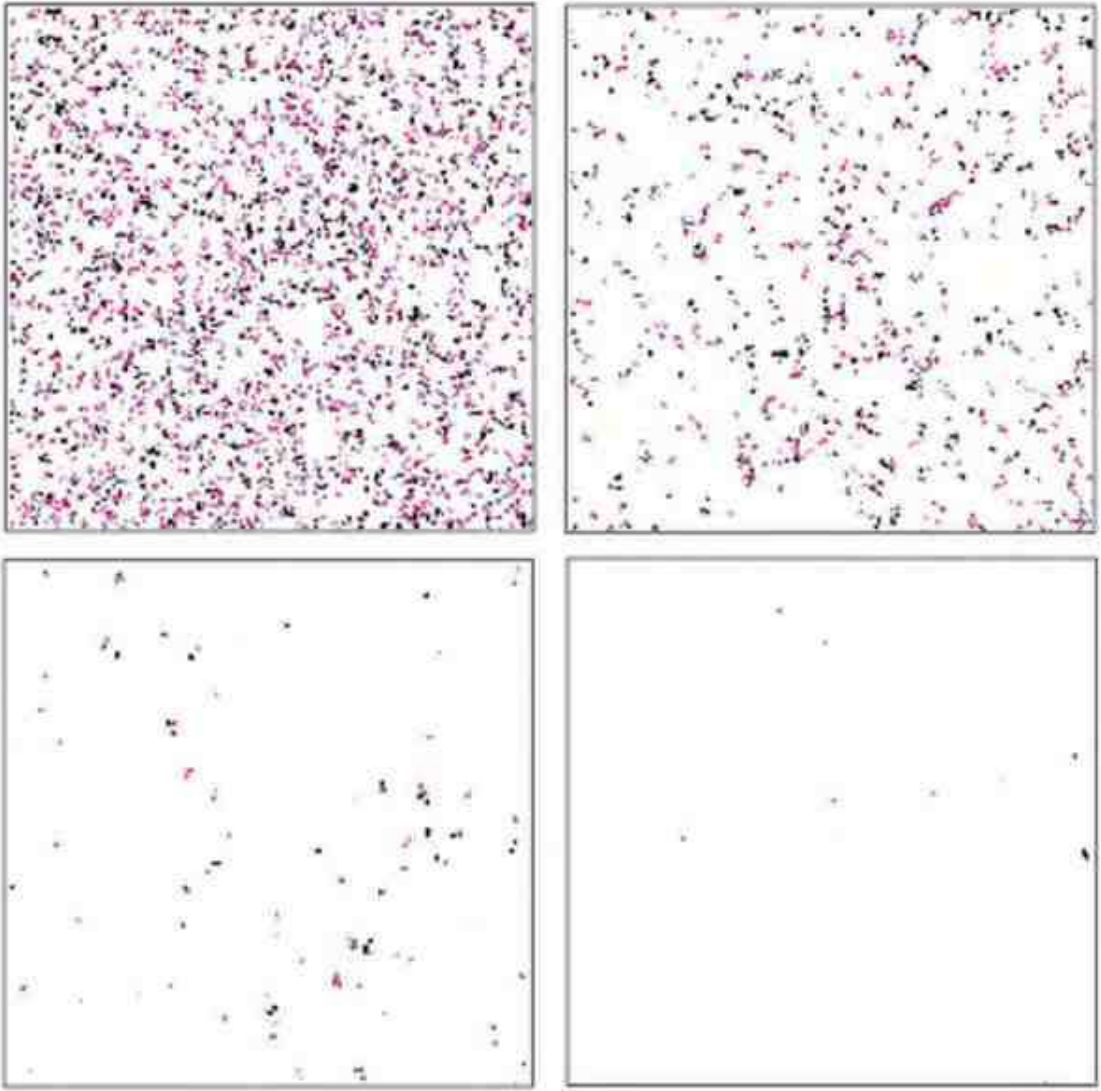}}
\caption{\label{fig: vortices}  Vortices in the $(x,y)$ plane
at different times: $t=2500$, $t=3250$, $t=5000$, $t=7500$.}
\end{figure}

\begin{figure}
\centerline{\includegraphics[width=0.4\textwidth]{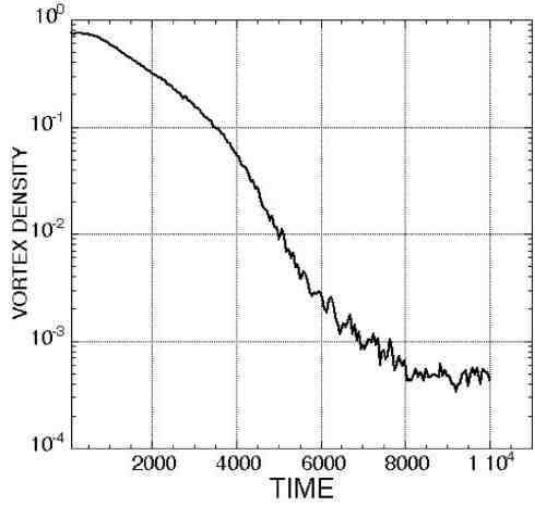}}
\caption{\label{fig: totvortices}  Evolution in time
of the density of vortices in a lin-log plot.}
\end{figure}

\begin{figure}

\centerline{\includegraphics[width=0.4\textwidth]{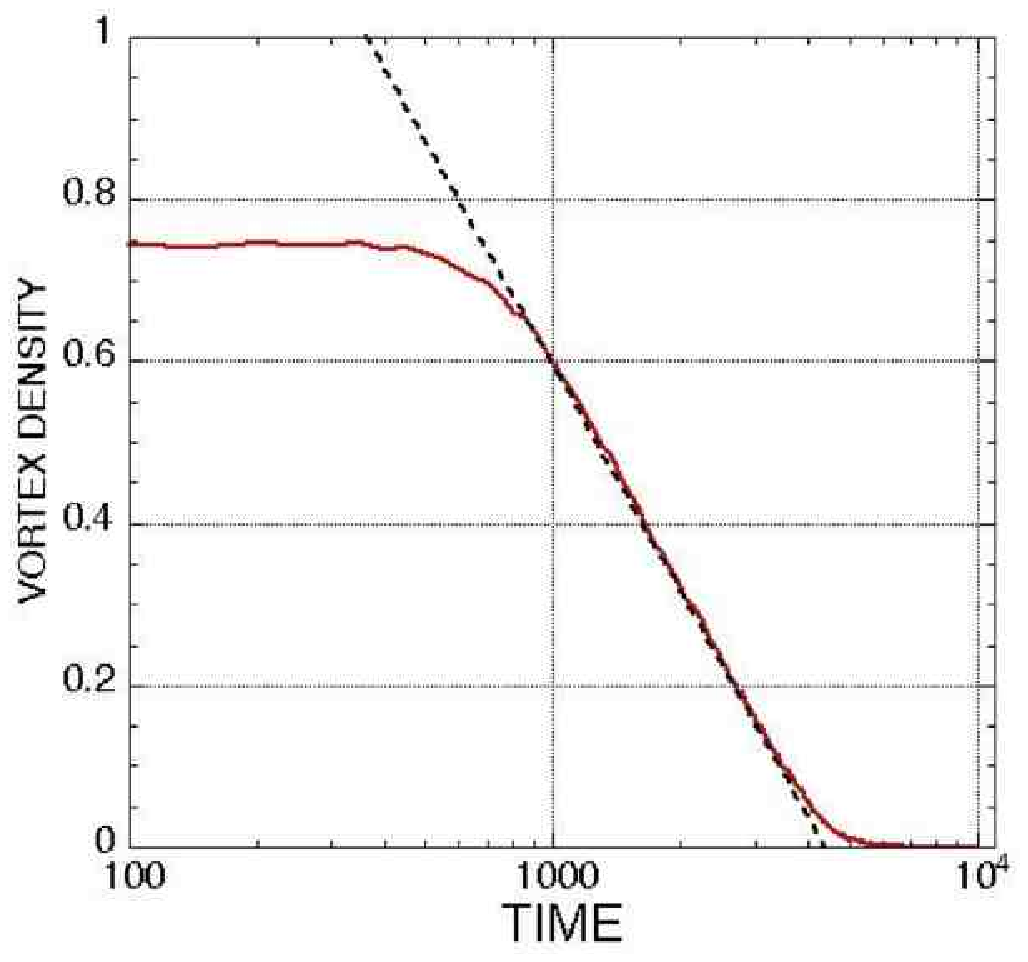}}
\caption{\label{fig:log-lin-vortices}  Evolution in time
of the density of vortices in a log-lin plot. 
The dashed line corresponds
to the fit $N_{vortices}=3.36-9.223 Log(t)$}
\end{figure}

On the other hand, field $|\Psi|$, shown in Figure \ref{fig: |psi|},
 still remains 
dominated by small-scale structure. In contrast with  $|\Psi|$,
field $\Psi$ contains an additional information -- the phase.
Thus, separation of the characteristic scales in Figure \ref{fig: repsi} and 
\ref{fig: |psi|} can be attributed to the fact that the phase correlation length
becomes much longer than the typical wavelength of sound (characterised by
fluctuations of   $|\Psi|$ as explained above in Section \ref{3-w-regime}).
This scale separation can also be seen by comparing the spectrum of
 $|\Psi|$, shown in Figure \ref{fig: spettro|psi|} with 
 the spectrum of $\Psi$ in Figures
\ref{fig: inverse01} and \ref{fig: inverse02}:
 one can see that the former is more flat
than the later.
Now that we have established that the phase is an important parameter, we
can measure its correlation length as the mean distance between the phase
defects -- vortices. Vortices in the GP model are points in which
$\Psi=0$. Some of such points correspond to the $2 \pi$ phase increment
when one goes once around them, whereas the other points gain  $-2 \pi$.
These vortices can be defined as positive and negative correspondingly.
In contrast with the Euler equation of the classical fluid, positive and
negative vortices can annihilate in the GP model and they can get
created ``from nothing''. Figure \ref{fig: vortices} shows a sequence of plates showing
the positive and negative vortex positions at several different 
moments of time. One can see that initially there were a lot of vortices,
which is not surprising because initial field is weak, i.e. close to zero
everywhere. However, at later times we see the number of vortices is
rapidly dropping, which means that the vortex annihilation process
dominates over the vortex-pair creations. 
The total number of vortices (normalised by $N^2$)
is shown as a function of time  in Figure 
\ref{fig: totvortices}, where one can see a 
fast decay.
The law of decay is best seen on the $log-lin$ plot, see
Fiigure \ref{fig:log-lin-vortices} where one can see
a regime 
\begin{equation}
N_{vortices} = A - B \log t
\label{nv}
\end{equation}
with $A=3.36$ and $B=0.9223$ which sets in at $t=800$ 
to  $t=3500$.\footnote{At present, we
  do not have a theoretical explanation of this law of decay.}
Thus, the phase correlation distance, being of the order of the
mean distance between the vortices, exhibits a fast growth in time.

\begin{figure}
\centerline{\includegraphics[width=0.4\textwidth]{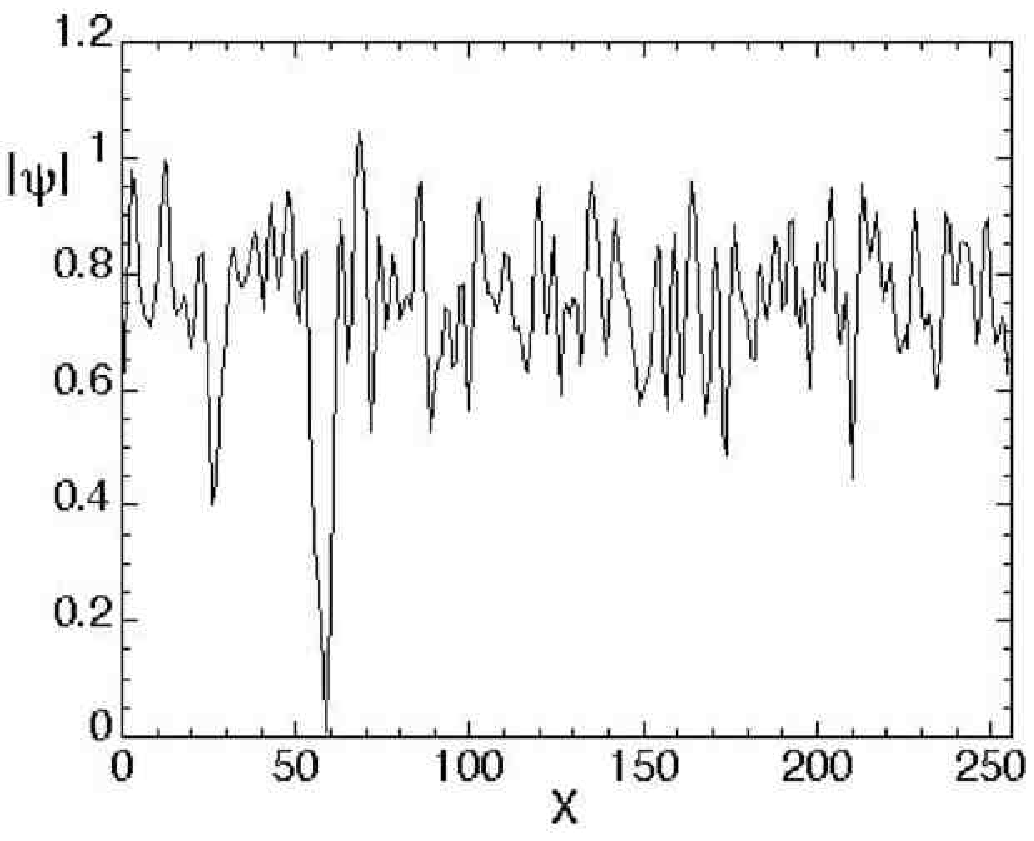}}
\caption{\label{fig: vortex} Slice of the field  $|\Psi|$ 
for constant $y$: a single vortex is visible in the plot.}
\end{figure}

\begin{figure}
\centerline{\includegraphics[width=0.4\textwidth]{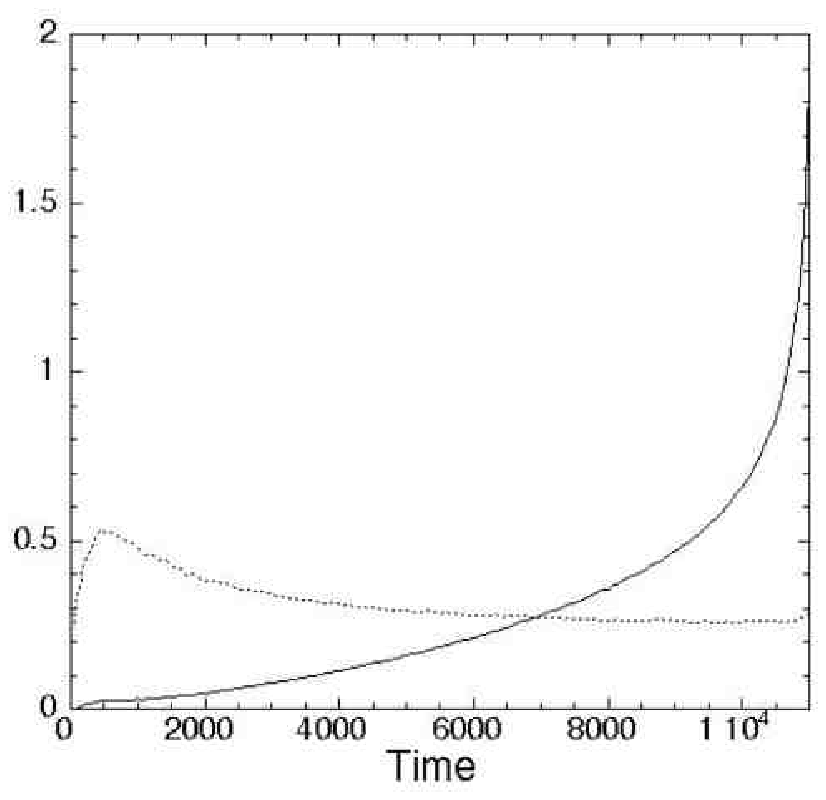}}
\caption{\label{fig: laplacian} The solid line represents the space-averaged
$|\nabla^2 \psi(x,y)|$; the dotted line is the space-averaged $|\psi(x,y)|^3$.
See text for comments.}
\end{figure}

Let us have a look at a slice of the field  $|\Psi|$ through
 typical vortices at a late
time when most of the them have annihilated, 
see Figure \ref{fig: vortex}.
One can see that  $|\Psi|$ is close to zero (i.e. both $Re[\psi]$ and
$Im[\psi]$ cross zero) at the vortex centres and that
it sharply grows to order-one values  (``heals'') at small distances
from the vortex centres which are much less than the distance
between the vortices. This means that these vortices represent
fully nonlinear coherent structures each of which can be approximately
seen as an isolated Pitaevski vortex solution \cite{Pitaevsky1961}.
In contrast, the initial vortices are too close to each other
to be coherent and they correspond to a nearly linear field.\footnote{
For this reason such vortices are sometimes called ``ghost
vortices''\cite{ghost}.} 
The moment when the mean inter-vortex 
separation becomes comparable to the healing length can be captured
by the intersection point of the graphs for the mean (space averaged)
nonlinear and the mean (space
averaged) Laplacian
terms in the GP equation, see Figure \ref{fig: laplacian}. 
This intersection (at $t=6950$) marks the
moment when mean nonlinearity becomes greater than the mean linear dispersion,
i.e. the Thomas-Fermi regime sets in.  
This regime could be thought of as the one
of a fully developed condensate when the nonlinearity, when measured
with respect to the zero level, is strong and therefore the 4-wave
WT description breaks down. However, as we will see in the
next section, we now have weakly nonlinear perturbations
if they are  measured with respect to a non-zero condensate
state. Evolution of such perturbations takes form
of three-wave acoustic turbulence.

\begin{figure}
\centerline{\includegraphics[width=0.4\textwidth]{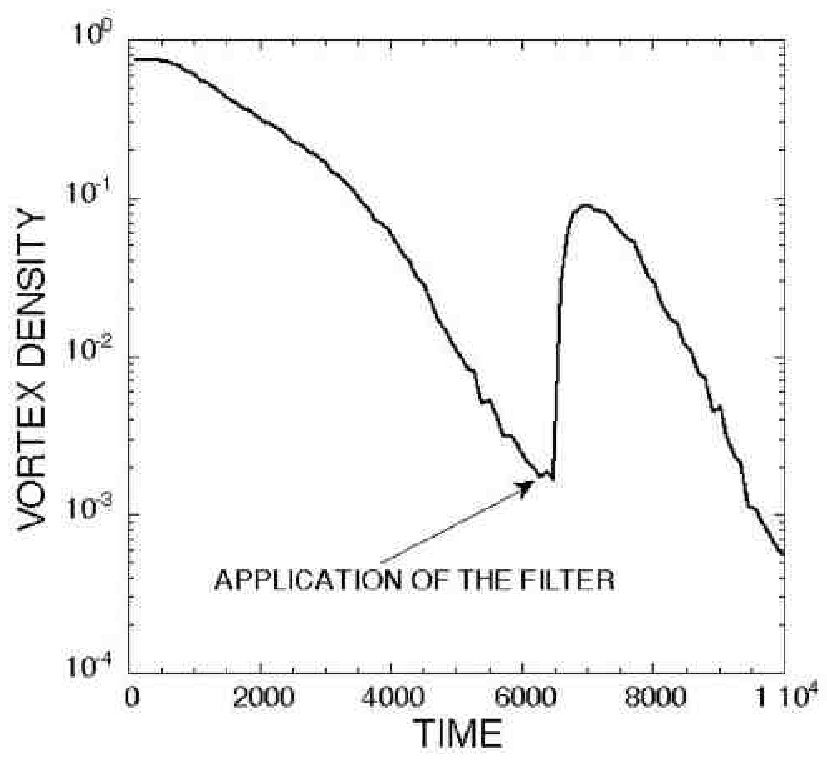}}
\caption{\label{fig: filter} Evolution of the vortex density in time. At time
$t=6500$, sound has been filtered according to the 
methodology described in the text.}
\end{figure}

What makes vortices annihilate? A positive-negative vortex pair,
when taken in isolation, would propagate with constant speed without
changing the distance between the vortices \cite{robertsjones}.
Thus, there should be an additional entity which could exchange
energy and momentum with the vortex pair and to allow them
annihilate. We note that the field   $|\Psi|$ is very ``choppy''
in the region between the vortices, see figure (\ref{fig: vortex}), 
and, therefore, it is natural to
conjecture that the missing entity is sound. To check this
conjecture, we perform the following numerical experiment.
At a desired time we filter the field and let it evolve further
without sound. The filtering is performed numerically in the 
following way: we have used a Gaussian filter in 
physical space and have smoothed the field around vortices.
The complex field $\psi$ is therefore convoluted with 
a normalised Gaussian function with standard deviation much 
smaller with respect to the mean distance between vortices. 
The filter is applied only in the region where no vortices are
located. 
The result of the filtering procedure on the evolution 
of the number of vortices is shown in Figure \ref{fig: filter}. 
We see that removing
the sound component does indeed reverse the vortex annihilation
process and for some time (until new sound gets generated 
from forcing) we observe
that the vortex creation process dominates.
We point out that the described above regime change,
accompanied by vortex annihilations, is very similar to the Kibble-Zurek
mechanism of the second-order phase transition in cosmology
\cite{kibble,zurek}.
This mechanism suggest that at an early inflation stage, Higgs
fields experience a symmetry breaking transition from ``false'' to
``true'' vacuum, and this transition is accompanied by a
reconnection-annihilation of ``cosmic strings'' which are 
3D analogs of the 2D point vortices considered in this paper.
To describe these fields, one normally uses nonlinear equations
of so-called Abbelian model \cite{peacock}, but the non-linear Klein-Gordon or
even the GP equation are sometimes used as simple models in cosmology
which retain similar physics \cite{peacock,zurekNLS}.

\subsubsection{Late condensation stage: acoustic turbulence}

It was predicted in \cite{DNPZ} that the turbulent condensation
in the GP model will lead to creation of a strong coherent
mode with $k=0$ such that the excitations at higher wave numbers
 would be weak compared to this mode. If this is the case,
one can expand the GP equation about the new equilibrium state,
uniform condensate, use Bogoljubov transform to find new normal
modes and a dispersion relation for them, equation (\ref{bogolubov}), and to 
obtain a new WKE for this system that would be characterised by
by three-wave interactions, (\ref{KE3wave}). 
However, as we saw
in Figure \ref{fig: inverse02}
the peak at small $k$ remains quite broad that is
the coherent condensate, if present, remains somewhat non-uniform.
Despite of this non-uniformity, 
one can still use the approach of  \cite{DNPZ}
if there is a scale  separation between the condensate
coherence length (intervortex distance) and the sound wavelength
and if the sound amplitude is much smaller than the one of the condensate.

\begin{figure}
\centerline{\includegraphics[width=0.4\textwidth]{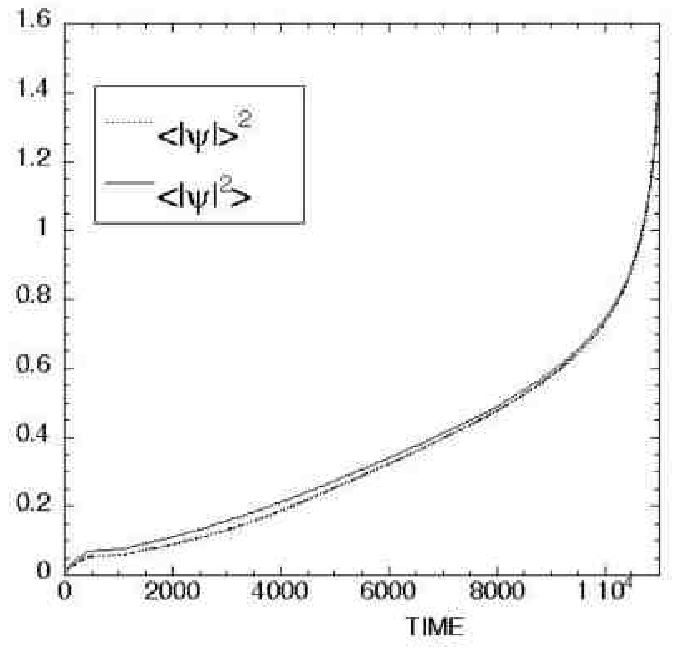}}
\caption{\label{fig:psi-graph} 
Evolution in time of $<|\Psi|^2>$ and  $<|\Psi|>^2$. }
\end{figure}

\begin{figure}
\centerline{\includegraphics[width=0.4\textwidth]{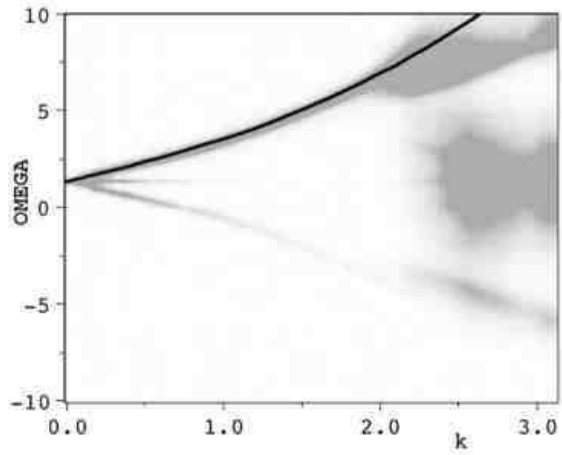}}
\caption{\label{fig:bogol_disp} Dispersion relation 
calculated from numerical simulation compared
with the upper branch of the Bogoljubov dispersion relation (solid line)}
\end{figure}

We have already seen tendency to the scale separation in
figures \ref{fig: inverse02}, \ref{fig: repsi},
\ref{fig: |psi|}, \ref{fig: spettro|psi|}.
On the other hand, smallness of the sound intensity
can be seen in figure \ref{fig:psi-graph} which compares (space-averaged)
 $<|\Psi|^2>$ and  $<|\Psi|>^2$. We see that at the late stages
these quantities have very close values which means that
the deviations of  $|\Psi|$ from its mean value (condensate) are weak.
Thus, both conditions for the weak acoustic turbulence to exist
are satisfied at the late stages.
However, the best way to check if the
condensate perturbations
do behave like weakly nonlinear
sound waves obeying the Bogoljubov dispersion relation
consists in plotting the square of the absolute value of the space-time Fourier transform of $\Psi$.
This result is given in Figure \ref{fig:bogol_disp}
for the latest stage of the simulation (from time
$t=10488$ to $t=11000$).  Note that for each $k$ the spectrum
has been divided by its maximum in order to be able to follow the 
dispersion relation up to high wave numbers.

The normal variable for the Bogoljubov sound is given
in terms of $\psi$ by expressions (\ref{psi_phi}) and  (\ref{bogolubov}) ,
and, therefore, when plotting the
Bogoljubov dispersion  (\ref{bog_disp}), we should add a constant frequency
of the condensate oscillations, $\omega_0 = \langle|\Psi|^2 \rangle$.
 One can see that the main branch of the spectrum does
follow the Bogoljubov law up to the wave numbers which correspond to
the dissipation range.\footnote{At the same time, these wave numbers
are of order of the inverse healing length, and it is unclear whether
the Bogoljubov mode is not seen there due to the wave dissipation or due to
contamination of this range by the broadband (in frequency) vortex motions.}
Further, the wave distribution is quite narrowly concentrated
around the Bogoljubov curve which indicates that these
waves are weakly nonlinear. 
Note that the lower branch in Figure \ref{fig:bogol_disp}
is related to the $\bar a_k$ contribution to expression
(\ref{bogolubov}) which vanishes at larger $k$.
Importanlty, we can also see  the middle (horizontal) branch
with frequency   $\omega_0$ which quickly fades away at finite $k$'s
and which corresponds to the coherent large-scale condensate component.

\begin{figure}
\centerline{\includegraphics[width=0.4\textwidth]{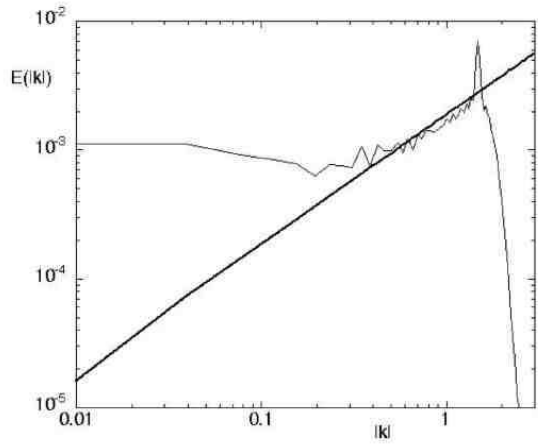}}
\caption{\label{fig:E} $E(k)=E(k_1)+E(k_2)$   at the latest stage of the simulation  (see equation (\ref{hamilt})) }
\end{figure}

Now let us consider the energy spectrum. The GP Hamiltonian
can be written in terms of both real and Fourier quantities,
\begin{equation}
H = \int ( |\nabla \Psi |^2 + {1 \over 2} |\Psi|^4) \, d{\bf x} =
\int (k^2 |\hat \Psi |^2 + {1 \over 2} | \hat{ \rho} |^2 ) \,  d{\bf k}, 
\label{hamilt}
\end{equation}
where $ \rho = |\Psi |^2. $
Thus, we measure the 1D energy spectrum in this case
as $E(k) = E_1(k) + E_2(k) $ with $E_1(k) = k^3 |\hat \Psi |^2 $
and $E_2(k) = {k \over 2} | \hat{ \rho} |^2 $
\begin{figure}
\centerline{\includegraphics[width=0.4\textwidth]{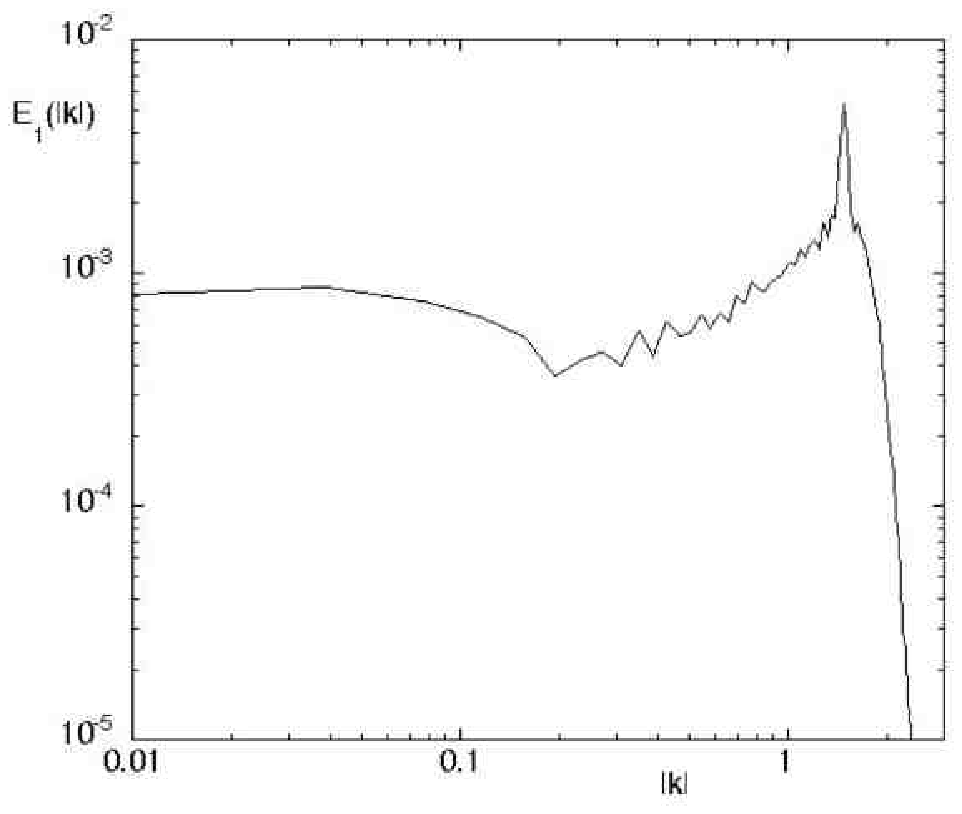}}
\caption{\label{fig:E1} $E_1(k)$  at the latest stage of the simulation 
(see equation (\ref{hamilt})).}
\end{figure}
\begin{figure}
\centerline{\includegraphics[width=0.4\textwidth]{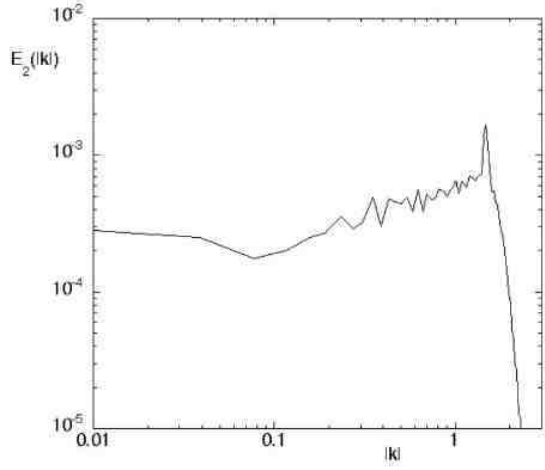}}
\caption{\label{fig:E2} $E_2(k)$  at the latest stage of the simulation
  (see equation (\ref{hamilt})).}
\end{figure}
%
The contributions to the energy spectrum $E_1$ and $E_2$ as well
as the total spectrum $E(k)$ at a time corresponding to the
acoustic regime are shown in figures \ref{fig:E1},
\ref{fig:E2} and \ref{fig:E} respectively.
%
We  see that at small scales the total energy spectrum
$E(k)$ scales as $1/k$ which is a thermodynamic energy
equipartition solution in this case.

\section{Conclusions}
Firstly, we confirmed WT predictions of the energy spectra
in the down-scale and up-scale inertial intervals in the
cases when the fluxes are absorbed by dissipation in the
end of the inertial interval (so that no condensation or
build-up is happening). In both of these cases we observed
spectra with an exponent corresponding to the energy
equipartition thermodynamic solution $N_k \sim 1/k$ (which formally
coincides with the exponent for the energy cascade solution).
By looking at the shape of the frequency-wave number 
mode distributions, we verified that the turbulence is weak.

Secondly, we studied a system without dissipation at
large scales. We observed a process of Bose-Einstein condensation
and formation of a coherent large-scale mode which happens
via annihilating vortices. This scenario is similar to the
Kibble-Zurek phase transition in cosmology which involves
annihilating cosmic strings.
We established that the process of the vortex annihilation 
is due to the presence of sound.
The condensate correlation length, which in our case is of the
order of the mean inter-vortex distance, turns into infinity
in a finite time as $\lambda \sim 1/(\log t^* - \log t)^{1/2}$, 
see equation (\ref{nv}).

We confirmed numerically that in late condensation stages
the system can be described as a weakly nonlinear acoustic
turbulence on the background of a quasi-uniform coherent
condensate. Namely, we confirmed that the wave excitations
are narrowly distributed around the Bogoljubov dispersion
law, i.e. that the turbulence is (i) acoustic and (ii) weak.
We observed a spectrum that corresponds to the energy
equipartition solution of the 3-wave kinetic equation
for such acoustic turbulence. 

An interesting question to be addressed in future is to
what extend the findings of this work are relevant to the
3D GP model. We can speculate that the energy spectra may
have a different nature in 3D and, in particular, may expect
formation of the Kolmogorov-like spectra corresponding
to the energy and the particle cascades. On the other hand,
it is reasonable to expect that the Kibble-Zurek scenario
of condensation will persist in the 3D case, i.e. the
correlation length will grow because of the reconnecting 
and shrinking vortex loops. It is also likely that such
vortex loop shrinking will be facilitated by the sound component.
Computations of 3D GP equation in a non-turbulent setting
were done in \cite{berloff} where such processes as vortex
reconnection and the role of the acoustic component were
considered. Turbulent setting will be more taxing on the 
computing resources
due to the great variety of scales involved and, therefore, necessity
of high resolution and long computation times. \\

\noindent {\bf Acknowledgments} \\
We thank Al Osborne for discussions in the early stages of the work.

\end{document}